\documentclass[submission,copyright,creativecommons]{eptcs}
\providecommand{\event}{PLACES2017} 
\usepackage{breakurl}             

\usepackage[nounderscore]{syntax}
\usepackage{amsmath}
\usepackage{amssymb}
\usepackage{array}
\usepackage{tabularx}
\usepackage{pdfpages}
\usepackage{amssymb}
\usepackage{amsthm}

\newcommand{\ignore}[1]{ }

\newcommand{\trans}[2]{\xrightarrow[#2]{#1}}
\newcommand{\arrow}{\rightarrow}

\newtheorem{proposition}{Proposition}
\newtheorem{definition}{Definition}
\newtheorem{theorem}{Theorem}

\makeatletter
\def\overbracket#1{\mathop{\vbox{\ialign{##\crcr\noalign{\kern3\p@}
      \downbracketfill\crcr\noalign{\kern3\p@\nointerlineskip}
      $\hfil\displaystyle{#1}\hfil$\crcr}}}\limits}
\def\underbracket#1{\mathop{\vtop{\ialign{##\crcr
      $\hfil\displaystyle{#1}\hfil$\crcr\noalign{\kern3\p@\nointerlineskip}
      \upbracketfill\crcr\noalign{\kern3\p@}}}}\limits}
\def\overparenthesis#1{\mathop{\vbox{\ialign{##\crcr\noalign{\kern3\p@}
      \downparenthfill\crcr\noalign{\kern3\p@\nointerlineskip}
      $\hfil\displaystyle{#1}\hfil$\crcr}}}\limits}
\def\underparenthesis#1{\mathop{\vtop{\ialign{##\crcr
      $\hfil\displaystyle{#1}\hfil$\crcr\noalign{\kern3\p@\nointerlineskip}
      \upparenthfill\crcr\noalign{\kern3\p@}}}}\limits}
\def\downparenthfill{$\m@th\braceld\leaders\vrule\hfill\bracerd$}
\def\upparenthfill{$\m@th\bracelu\leaders\vrule\hfill\braceru$}
\def\upbracketfill{$\m@th\makesm@sh{\llap{\vrule\@height3\p@\@width.7\p@}}%
  \leaders\vrule\@height.7\p@\hfill
  \makesm@sh{\rlap{\vrule\@height3\p@\@width.7\p@}}$}
\def\downbracketfill{$\m@th
  \makesm@sh{\llap{\vrule\@height.7\p@\@depth2.3\p@\@width.7\p@}}%
  \leaders\vrule\@height.7\p@\hfill
  \makesm@sh{\rlap{\vrule\@height.7\p@\@depth2.3\p@\@width.7\p@}}$}
\makeatother
\newcommand{\reduct}[1]{\raise3.5pt\hbox{\scriptsize$\ulcorner$}\kern-3.0pt#1\kern-2.6pt\raise3.5pt\hbox{\scriptsize$\urcorner$}}

\makeatletter
\newcommand{\rd}[1][x]{%
  \@ifnextchar[{\rd@ii[#1]}{\rd@ii[#1][v]}
}
\def\rd@ii[#1][#2]{%
  \@ifnextchar[{\rd@iii[#1][#2]}{\rd@iii[#1][#2][]}
}
\def\rd@iii[#1][#2][#3]{%
  {\texttt{rd{#3}}_{#1}^{#2}}
}
\makeatother

\makeatletter
\renewcommand{\wr}[1][x]{%
  \@ifnextchar[{\wr@ii[#1]}{\wr@ii[#1][v]}
}
\def\wr@ii[#1][#2]{%
  {\texttt{wr}_{#1}^{#2}}
}
\makeatother

\makeatletter
\newcommand{\srd}[1][x]{%
  \@ifnextchar[{\srd@ii[#1]}{\srd@ii[#1][v]}%
}
\def\srd@ii[#1][#2]{%
  {\texttt{srd}_{#1}^{#2}}
}
\makeatother

\makeatletter
\newcommand{\swr}[1][x]{%
  \@ifnextchar[{\swr@ii[#1]}{\swr@ii[#1][v]}
}
\def\swr@ii[#1][#2]{%
  {\texttt{swr}_{#1}^{#2}}
}
\makeatother

\newcommand{\R}{\ensuremath{\mathbb{R}}}
\newcommand{\Rinf}{\ensuremath{\mathbb{R}^{\infty}}}

\title{Best-by-Simulations:  A Framework for Comparing Efficiency of Reconfigurable Multicore 
Architectures on Workloads with Deadlines}
\author{Sanjiva Prasad
\institute{Indian Institute of Technology Delhi\\ New Delhi, India}
\email{sanjiva@cse.iitd.ac.in}
}
\def\titlerunning{Best-by-Simulations}
\def\authorrunning{S. Prasad}
\begin{document}
\maketitle

\begin{abstract}
Energy consumption is a major concern in multicore systems. 
Perhaps the simplest strategy for reducing energy costs is to use only as many cores as necessary while still being able to deliver a desired quality of service.
Motivated by earlier work on a dynamic (heterogeneous) core allocation scheme for H.264 video decoding that reduces energy costs while delivering desired frame rates, we formulate operationally the general problem of executing a sequence of actions on a reconfigurable machine while meeting a corresponding sequence of absolute deadlines, with the objective of reducing cost.
Using a transition system framework that associates costs (e.g., time, energy) with executing an action on a particular resource configuration, we use the notion of amortised cost to formulate  in terms of simulation relations appropriate notions for comparing deadline-conformant executions.
We believe these notions can provide the basis for an operational theory of optimal cost executions and performance guarantees for approximate solutions, in particular relating the notion of simulation from transition systems to that of competitive analysis used for, e.g., online algorithms.

\end{abstract}

\section{Introduction}\label{INTRODUCTION}

Video decoding \cite{VIDEO}, an almost ubiquitous application on machines ranging from mobile phones to server machines, is amenable to execution on \textit{embedded multicore} platforms --- multi-threaded  implementations of the H.264 codec \cite{H264}  run on processors such as Intel Silvermont (homogeneous multicore) \cite{silvermont} and ARM Cortex A15 (heterogeneous
multicore, based on the delightfully named big.LITTLE architectural model). 
High video quality means better resolution and higher frame rates, which in turn requires more computation and thus more energy.
The required frame rate determines a \textit{budgeted per-frame decode time}, and thus a \textit{series of deadlines} for decoding each of a series of frames. 
The standard implementations utilise as many cores as available on the multicore platform in order to meet performance requirements.

To reduce energy consumption, Pal \textit{et al.} proposed and implemented dynamic core allocation schemes in which cores are switched on or off using clock gating (or in heterogeneous multicores, smaller cores used instead of larger ones) according to the per-frame decoding requirements \cite{PalSPP16}. 
The basic idea is that since frames are often decoded well within the budgeted decode time, if deadlines can still be met by using fewer/smaller cores for decoding a frame, then the same performance can be achieved with lower energy consumption.
By measuring “slack” and “overshoot” over the budgeted decode time and amortising these across frames, their schemes are able to save energy without missing any performance deadlines. 
Simulations on Sniper \cite{SNIPER} for timing and McPAT \cite{MCPAT} for energy measurements show energy savings of 6\% to 61\% while strictly adhering to the required performance of 75 fps on homogeneous multicore architectures, and 2\% to 46\% while meeting a performance of 25 fps on heterogeneous multicore architectures.

There, however, is no corresponding theoretical framework for (1) justifying the correctness of such schemes, or (2) comparing the performance of difference multicore (re)configurations on a given workload. 
While there are algorithmic optimisation approaches for  structured problems in which the trade-offs between achieving an objective in a timely manner and the cost incurred for doing so are expressed, there are few formulations in \textit{operational semantic} terms. 

In this paper, we generalise the video decoding problem to the following abstract setting: ``Suppose we are given a workload consisting of a sequence of actions, each of which has to be performed by a given deadline. 
Suppose there are different computational machine configurations (let  $\mathit{Conf}$ denote the set of these configurations) on which these actions may be executed, with possibly different costs per action-configuration combination.
\begin{enumerate}
	\item Can the sequence of actions  be performed on some machine configuration while meeting each deadline?
    \item Is a given reconfiguration scheme (strategy/heuristic) correct (i.e., meets all deadlines)?
    \item How can we compare  the cost of execution according to a reconfiguration heuristic/strategy versus that on the baseline configuration?
    \item Is it possible to express performance guarantees of a reconfiguration scheme with respect to an optimal strategy?
\end{enumerate}
This generalisation allows us to examine the execution of arbitrary programs, expressed as a sequence of atomic tasks or workloads (not just video decoding) on a variety of architectures (not only multicores), particularly those that support reconfiguration, where we seek to reduce the cost of execution (not merely energy), subject to some performance deadlines.  

The trade-offs involved are non-trivial, since different actions require differing processing times, with there being no simple method for anticipating the number and kinds of future actions (the problem is posed as an ``online'' one). 
For example, it is not entirely obvious whether while trying to save energy by using a slower computational configuration to perform an action, we will have enough time for processing subsequent actions without missing deadlines.
On the other hand, being too conservative and operating only on the fastest configurations may mean forgoing opportunities for saving energy.
Note that the problem is not of task scheduling but rather of resource allocation to meet a performance constraint (and then of finding close-to-optimal-cost executions; also see \S\ref{RELATED}).

In this work, we present an operational semantics framework for specifying the execution of a workload in terms of \textit{cumulative weighted transition systems}, which lets us record execution times (and then energy consumption).  
We then use the notion of simulation to express the execution of fixed workloads on different computational resource configurations as well as the specification of a deadline-meeting execution (\S\ref{BYSIM}).   
An important feature of our framework is that it is not confined to dealing with finite-state systems and finite workloads, and so applies to both finite and infinite runs of a system. 
In \S\ref{CAPABILITY}, we compare the capabilities of different resource configurations in executing a specified workload, with Propositions \ref{PREORDER-EQUIV}--\ref{SUFFIX} providing some useful properties.
The framework is extended to deal with \textit{reconfiguration} (\S\ref{RECONFIG}), following which we show the correctness of the scheme proposed by Pal \textit{et al.} in \cite{PalSPP16} (Theorem \ref{CORRECTNESS}). 
The weighted transition systems are extended to account for energy consumption in \S\ref{ENERGY}, using which we are able to formally state that the scheme of Pal \textit{et al.} performs better than the baseline configuration
(Theorem \ref{BETTER}). 
The formulation allows us to examine an instance where there is a trade-off between efficiency in energy consumption versus satisfying timeliness constraints.
We continue in \S\ref{BEST} with a discussion on how one may formulate comparisons of performance with \textit{optimal executions}, and propose a notion of simulation with performance within a constant factor $c$.
We envisage this is the first step towards relating operational formulations of correctness with the \textit{competitive analysis} of approximation algorithmic schemes in the case of possibly infinite executions. 
\S\ref{CONCLUSIONS} mentions a possible application in security that illustrates how the framework can address problems that go beyond meeting
time  deadlines.
We then briefly discuss how the framework can be modified to deal with online scheduling of concurrently enabled threads during program execution on a reconfigurable machine. 
We conclude with a short statement on our future goals of developing further connections between operational notions such as simulation and approaches used in the analysis of relative and absolute performance guarantees of (online) algorithms.

\subsection{Related Work}\label{RELATED}

\textit{Timed automata} are the preferred operational framework for 
specifying time-related properties of systems. 
In particular, the cost-optimal reachability problem has been studied both  in a single-cost \cite{AlurTP01} and in multi-cost settings \cite{LarsenR08}.
Bouyer\textit{ et al.} have studied issues relating to minimising energy in infinite runs within the framework of weighted (priced) timed automata \cite{BouyerFLMS}. 
Specifically, they have examined  the  construction of
infinite schedules for \textit{finite} weighted automata and one-clock weighted
timed automata, subject to boundary constraints on the accumulated weight.
However, we are unaware  of an automata-based formulation of our general deadline-constrained execution problem, especially with respect to minimising cost (energy consumption), where the times/costs are cumulative and unbounded, i.e., where the state spaces and value domains (and possibly the alphabet) are \textit{not finite}.  

The seminal work in the use of process algebra for performance analysis is by Hermanns \textit{et al.} \cite{HermannsHK02}.
G\"{o}tz et al \cite{Gotz+:1993}  have used stochastic process algebra in studying correctness and performance analysis of multiprocessors and distributed system design.
Klin and Sassone \cite{Klin,KlinS} have explored using monoidal structures for stochastic SOS, an elegant approach that unifies various different operational semantic models into a single algebraic frame. 
This approach has  been taken further by Bernardo \textit{et al.} \cite{Bernardo} in finding a unifying structure for dealing with probabilistic, stochastic and time-dependent non-determinism.
The theory of weighted automata has been studied by Almagor, Kupferman and others \cite{AlmagorBK}.
Their weighted automata approach allows optimisation problems to be formulated as runs for finite words yielding values in a tropical semiring. 

The dynamic reconfiguration scheme we study may be transformed to an instance of \textit{dynamic speed scaling} in task scheduling \cite{Albers}, where  tasks have strict deadlines and a scheduler has to construct feasible schedules while minimising energy consumption.  
Instead of using multiple cores, dynamic speed scaling allows the speed of the processor to be changed,  assuming a model where power consumption increases exponentially with the speed of the processor ($P(s) = s^{\alpha}$).
The polynomial-time YDS algorithm \cite{YaoDS} finds optimal schedules in the offline case when all tasks and their requirements are known \textit{a priori} ($\mathcal{O}(n^3)$ for a naive implementation, which can be improved to $\mathcal{O}(n^2 log n)$).
The main idea is to find maximum density intervals, and schedule tasks occurring within them according to an earliest deadline first (EDF) policy.
Tasks may be left unexecuted, and may be pre-empted. 
On the one hand, YDS deals with the more general problem of task scheduling, but on the other hand assumes a given relationship between power and speed, unlike our formulation, which leaves this relationship un(der)specified.  
Results about the competitive analysis of online versions of the algorithm (Average Rate and Optimal Available) have been given\footnote{An online algorithm ALG  is called $c$-competitive if for every input task sequence, the objective function value of ALG  is within $c$  times the value of an optimal solution for that input.}, assuming the exponential power-speed relationship.
These bounds have been shown to be essentially tight \cite{BansalKP}.
Bansal \textit{et al.} have also used the concept of slack and urgency in a variant of the problem, where deadlines may be missed but throughput maximisation is the objective function, presenting an online algorithm that is 4-competitive \cite{BansalCLL}.

\section{Getting the Job Done: An Operational Model}\label{BYSIM}

\paragraph{Preliminaries.}  We define a weighted transition system, workloads, deadlines and executing a workload respecting deadlines.

\begin{definition}
A weighted transition system 
$\mathcal{T} = (Q, \mathcal{A}, \mathcal{W}, \xrightarrow{}, Q_0, O)$ consists of 
a set of states $Q$; 
an input alphabet $\mathcal{A}$; 
an output domain  $\mathcal{W}$;
 a cost-weighted transition relation $\_ \trans{\Box}{\Box} \_: Q \times \mathcal{A} \times Q  \times \mathcal{W}$; 
 a set of initial states $Q_0 \subseteq Q$; 
 and an observation function $O: Q \arrow  \mathcal{W}$.
\end{definition}
A weighted transition system is a minor modification of an input-output Moore-style transition system. The major difference is that instead of an
output set/alphabet we have a (monoidal) weight domain, and the transition relation, written $q \trans{a}{w} q'$, which maps a transition from $q$ on $a$ to $q'$  to a  weight $w \in \mathcal{W}$.  This may be thought of the combination of a transition relation $\Delta \subseteq  Q \times \mathcal{A} \times Q$ and a cost function $c: \Delta \arrow \mathcal{W}$. 
Further, we assume additional structure on the weight domain  --- (1) it is a
partially ordered set $\langle \mathcal{W},  \leq_{\mathcal{W}} \rangle$
(2) it is also a  \textit{monoid} $\langle \mathcal{W}, \oplus, \oldstylenums{0} \rangle$, where $\oldstylenums{0}$ is the identity element for $\oplus$. 
The operation $\oplus$ is monotone and expansive w.r.t.  $\leq_{\mathcal{W}}$, i.e., for all $x, y, z \in \mathcal{W}$,  $x \leq_{\mathcal{W}} y$ implies $x \oplus z \leq_{\mathcal{W}} y  \oplus z$, and $x \leq_{\mathcal{W}} x \oplus y$ and $y \leq_{\mathcal{W}} x \oplus y$.
For a finite sequence $a_1 \ldots a_n$, we define
$q_0 \trans{a_1 \ldots a_n}{w} q_n ~ = w = ~ \bigoplus_{i=1}^{n} w_i$ where
$q_{i-1} \trans{a_i}{w_i} q_{i} ~~(i \in \{1,\ldots, n\})$.  
When $n=0$,  the weight $w = \oldstylenums{0}$, and otherwise
$\bigoplus_{i=1}^{n} w_i ~=~  ( \ldots (\oldstylenums{0} \oplus w_1) \ldots \oplus w_n)$ --- the notation is unambiguous even if $\oplus$ is not commutative.
A weighted transition system is \textit{cumulative} if whenever  $q \trans{a}{w}q'$ then $O(q') = O(q)  \oplus w$ (and consequently, $O(q) \leq_{\mathcal{W}} O(q')$).
It is sometimes useful to extend $\mathcal{W}$ to contain a maximum and annihilating element $\omega$ for $\oplus$, i.e., $x \oplus \omega ~=~ \omega = \omega \oplus x$ and $ x \leq_{\mathcal{W}} \omega$ for all $x$.
We write  $q \trans{a}{-} q'$  if $q \trans{a}{w} q'$ for some weight $w <_{\mathcal{W}} \omega$, and so can write $q \trans{a}{\omega} q'$ whenever  $q \not\trans{a}{-} q'$.
For the motivating example, we will consider $\mathcal{W} = (\Rinf, +, 0)$ (with $\omega = \infty$), which allows us to model time and deadlines.

We recast the notion of simulation for weighted transition systems. 
Note that our formulation uses the observation function $O$ to compare weights.
\begin{definition}
	Suppose
	$\mathcal{T}_1 = (Q_1, \mathcal{A}, \mathcal{W}, \xrightarrow{}, Q_{1o}, O_1)$ and 
	$\mathcal{T}_2 = (Q_2, \mathcal{A}, \mathcal{W}, \xrightarrow{}, Q_{2o},
	 O_2)$ are weighted transition systems on the same input alphabet $\mathcal{A}$ and weight domain $\mathcal{W}$.
	 A simulation relation between $\mathcal{T}_1$ and $\mathcal{T}_2$ 
	  is a binary relation $R \subseteq Q_1 \times Q_2$ 
	  such that  $(p,q) \in R$ implies
	   (i) $O_2(q) \leq_{\mathcal{W}} O_1(p)$; 
	  and (ii) whenever $p \trans{a}{-} p'$, there exists $q'$ such that $q \trans{a}{-} q'$ and $(p',q') \in R$.
\end{definition}
We say $q$ simulates $p$ if $(p,q)$ is in \textit{some} simulation.
Transition system $\mathcal{T}_2$ simulates $\mathcal{T}_1$ if for \textit{all} $p \in Q_{1o}$ there is a $q \in Q_{2o}$ such that $q$ simulates $p$. 
That is, from $q$ one can do everything that the other can from $p$, and with a lower weight.  

\begin{proposition}\label{WEIGHT-SIM}
	Simulation relations include identity and are closed under composition and unions:
	(i) The identity relation $\{ (p,p) ~|~ p \in Q \}$ is a (weighted) simulation;
	(ii) If  $R_1$ and $R_2$ are weighted simulations, then so is $R_1 \circ R_2$.
	(iii) If $R_i ~~(i \in I)$ are simulation relations, then so is $\bigcup_{i \in I}~R_i$.
\end{proposition}
The largest simulation relation is thus a quasi-order (reflexive and transitive).

\paragraph{Workloads with Deadlines.} \ \ 
A workload is a (finite or infinite) sequence $\mathbf{a}= a_1 a_2 \ldots$, such that each $a_i \in \mathcal{A}$.  
Suppose with each $a_i$, we have a corresponding \textit{budgeted time} $b_i \in \R$.  
Assume that the \textit{actual time} taken to perform each task $a_i$ on a
machine configuration $r \in \mathit{Conf}$ is given by $\tau(r,a_i) = t_i$.
For simplicity, we assume $\tau$ is a function, though in practice the same computational task $a_i$ may take differing amounts of time under different circumstances (e.g., ambient temperature, memory resources consumed by other tasks, etc.).

A na\"{i}ve formulation of being able to satisfy this workload on configuration $r$ is that $\forall i, 0 \leq i: t_i \leq b_i$, i.e., the actual time taken for each frame is less than the budgeted time.  
For frame-decoding, the budgeted time is the inverse of the desired frame rate.
However, this is overly conservative, since it does not allow for the fact that one can begin processing the next frame early,  thus amortising across frames using the slack earned by decoding a frame well within its budgeted time to offset overshoot incurred when taking longer than the budgeted time to decode another frame.
Therefore, we consider a cumulative formulation, choosing to model
a workload $\mathbf{a}$ together with a corresponding sequence of \textit{absolute deadlines} $\mathbf{d} = d_1 d_2 \ldots$, where $d_i = \Sigma_{j \leq i}~ b_j$.

We can \textit{specify} a workload $\mathbf{a} = a_1 \ldots$ with corresponding deadlines $\mathbf{d} = d_1 \ldots$ as a \textit{deterministic} transition system $\mathbf{Spec}$ as $0 \trans{a_1}{b_1} d_1 \ldots d_{i-1} \trans{a_i}{b_i} d_i \ldots$, with $Q \subset \R$ and $O(d_i) = d_i$, where the $b_i$'s are the budgeted times for each action. 

The transition system $\mathcal{T}^{\mathbf{a}}_r$ for executing workload $\mathbf{a}$ on a machine configuration $r$ can be modelled in terms of ($r$ paired with) the cumulative time taken so far, i.e., $Q \subset \mathit{Conf} \times \R$, $O(\langle r, t \rangle)  = t$, and $\langle r, t \rangle \trans{a}{w}
\langle r, t' \rangle$ if $t' = \tau(r, a) + t$.  The initial state is $\langle r, 0 \rangle$. Note that for a given workload this also is a \textit{deterministic} transition system, i.e., a \textit{path}.

\begin{definition}
	We say that execution on a machine configuration $r$ ``by-simulates'' a specified workload $\mathbf{Spec}$ ($\mathbf{a}$ with corresponding deadlines $\mathbf{d}$) if there is a simulation relation between $\mathbf{Spec}$ and $\mathcal{T}^{\mathbf{a}}_r$ for this workload. 
\end{definition}
That is, the execution sequence on machine $r$ meets each deadline. 
The machine configuration $r$ is then said to be \textit{capable} of executing the specified workload with the expected quality of service; otherwise this configuration is incapable of doing so.  

\subsection{Good Enough: Comparing Configurations  Based on Capability}\label{CAPABILITY}

Consider a workload specification $\mathbf{Spec}$ (action sequence $\mathbf{a}$ with corresponding deadlines $\mathbf{d}$)
and  two computational resource configurations $r$ and $r'$.
We say that $r$ is \textit{at least as capable as} $r'$ in performing 
 $\mathbf{Spec}$, written $r' \preceq_{\mathbf{Spec}} r$,  
 if  $\mathcal{T}^{\mathbf{a}}_{r'}$ can by-simulate $\mathbf{Spec}$ implies that so can  $\mathcal{T}^{\mathbf{a}}_r$.
We say that $r$ and $r'$  are \textit{equi-capable} in performing $\mathbf{Spec}$, written $r \sim_{\mathbf{Spec}} r'$ if 
$\mathcal{T}^{\mathbf{a}}_{r'}$ by-simulates  $\mathbf{Spec}$ if and  only if   $\mathcal{T}^{\mathbf{a}}_r$ does. 
In other words, both resource configurations are capable of meeting the sequence of deadlines. 

\begin{proposition}\label{PREORDER-EQUIV}
	For every  workload $\mathbf{Spec}$, the relation  $\preceq_{\mathbf{Spec}}$ is a preorder, and $\sim_{\mathbf{Spec}}$ an equivalence. 
\end{proposition}

Without any additional conditions, we cannot say much about the relationship between the capabilities of different computational resources on \textit{different} workloads.   
Note that it is  possible for $r \sim_{\mathbf{Spec}} r'$ for some workload $\mathbf{Spec}$ but $r \not\sim_{\mathbf{Spec}'} r'$ for some other workload $\mathbf{Spec}'$.
We say that $r$ is \textit{elementarily at least as capable as} $r'$ if for each possible action $a:   \tau(r,a) \leq \tau(r',a)$.
\begin{proposition}\label{ELEMENTARY-PREORDER}
	If $r$ is \textit{elementarily at least as capable as} $r'$, then for any workload $\mathbf{Spec}$,  $r' \preceq_{\mathbf{Spec}} r$.
	\end{proposition}
This notion captures the intuition that the capability of a resource configuration is an inherent property (e.g., its speed) rather than peculiarly dependent on the action to be executed.  The following proposition relate capability with sub-sequences of actions. 

Capability and equi-capability are  prefix-closed (Proposition \ref{PREFIX}) and the notions
also  suffix-compose (Proposition \ref{SUFFIX}).
\begin{proposition}\label{PREFIX}
  Let $\mathbf{Spec}$ be a workload.
If $r' \preceq_{\mathbf{Spec}} r$ (respectively $r \sim_{\mathbf{Spec}} r'$) then for each prefix  $\mathbf{Spec}'$ of  $\mathbf{Spec}$,
 $r' \preceq_{\mathbf{Spec}'} r$ (respectively $r \sim_{\mathbf{Spec}'} r'$).
\end{proposition}

\begin{proposition}\label{SUFFIX}
	Let $\mathbf{Spec}$ be a finite workload of actions $a_1,\ldots,a_m$ with deadlines $d_1, \ldots, d_m$   and $\mathbf{Spec}'$ be another (finite or infinite) workload of actions $a'_1, \ldots a'_j, \ldots$, with deadlines $d'_1, \ldots, d'_j \ldots $.  
	Consider the sequenced workload
	 $\mathbf{Spec}'' = a_1, \ldots, a_m, a'_1, \ldots, a'_j, \ldots$, with deadlines $d_1, \ldots, d_m, d'_1+d_m, \ldots, d'_j+d_m, \ldots$. 
	 Then, if $r' \preceq_{\mathbf{Spec}} r$ and $r' \preceq_{\mathbf{Spec}'} r$
	 (respectively $r \sim_{\mathbf{Spec}} r'$ and $r \sim_{\mathbf{Spec}'} r'$), then 
	 $r' \preceq_{\mathbf{Spec}''} r$ (respectively $r \sim_{\mathbf{Spec}''} r'$).
\end{proposition}
In particular, if $r' \preceq_{\mathbf{Spec}} r$ (respectively $r \sim_{\mathbf{Spec}} r'$), then for any workload $\mathbf{Spec}'$ of which
$\mathbf{Spec}$ is a prefix, $r' \preceq_{\mathbf{Spec}'} r$ (respectively $r \sim_{\mathbf{Spec}'} r'$).

Note however that if $\mathbf{Spec}'' = a_1, \ldots, a_m, a'_1, \ldots, a'_n$, with deadlines $d_1, \ldots, d_m, d'_1+d_m, \ldots, d'_n+d_m$, 
and $r' \preceq_{\mathbf{Spec}''} r$ (respectively $r \sim_{\mathbf{Spec}''} r'$), while by Proposition \ref{PREFIX}, for 
$\mathbf{Spec} = a_1, \ldots, a_m$, with deadlines $d_1, \ldots, d_m$ we necessarily have
$r' \preceq_{\mathbf{Spec}} r$ (respectively $r \sim_{\mathbf{Spec}} r'$), it may \textit{not} be the case that for
$\mathbf{Spec}' = a'_1, \ldots, a'_n$, with deadlines $d'_1, \ldots, d'_n$, that we will have $r' \preceq_{\mathbf{Spec}'} r$ (respectively $r \sim_{\mathbf{Spec}'} r'$), because the resource configurations are capable of performing the latter part of the workload before the specified deadlines \textit{only} because of ``credit'' earned by completing the prefix $\mathbf{Spec}$ \textit{sufficiently early}.  

\paragraph{Resource Lattice.} 
In the video decoding applications, we assume that we have the \textit{elementarity} property,  based on the  assumptions made by Pal \textit{et al}:
(0) Decoding a frame can be cleanly decomposed into decoding of independent slices/macroblocks, assigned to different cores. 
(1) The decoding time for a frame is monotone non-increasing in the number of cores;
(2) In heterogeneous architectures, decoding time for a frame does not increase when moving from a small core to a big core.
Therefore, we can assume a lattice structure with ordering $\preceq$ applicable to \textit{any} workload $\mathbf{Spec}$, with a \textit{maximal} resource configuration $r_{max}$ being the one where all cores of all kinds are given work, and a minimal configuration $r_{min}$ which is one in which all cores are switched off (of course, not much happens on that minimal configuration).

In the sequel, we will only consider workloads $\mathbf{Spec}$ which \textit{can} be successfully executed (meeting all deadlines) on the maximal resource configuration $r_{max}$.  This will be considered the baseline configuration.

\subsection{Reconfiguration}\label{RECONFIG}
Till now we have considered only deterministic transition systems (paths) that arise for a given workload on a given configuration, and have compared  different configurations on their ability to handle a given workload.  
We now consider \textit{reconfigurable machines}.
Let  $\delta_{r,r'}$ denote the cost of changing configuration from $r$ to $r'$, with $\delta_{r,r}$ being 0. 
For simplicity, we assume any change of configuration to have a constant cost $\delta$.  
We can now define \textit{reconfigurable} execution to be the \textit{non-deterministic} transition system $\mathcal{N}$, obtained by modifying the earlier weighted transitions as follows:  $\langle r, t \rangle \trans{a}{w}
\langle r', t' \rangle$ if $t' = t + \delta_{r,r'} + \tau(r', a)$, denoting the cost of
changing configuration to $r'$ and then executing $a$.  
$w = \delta_{r,r'} + \tau(r', a)$.
The start state is $\langle r_{max}, 0 \rangle$.
The branching structure captures the various possibilities in choosing to reconfigure the machine at any stage in the execution. 

A reconfiguration scheme (algorithm/heuristic) defines a sub-transition system (a pruning) $\mathcal{T}$ of $\mathcal{N}$.
In general, this may be a non-deterministic transition system, embodying the possibility of reconfiguration according to the scheme, which is why we use simulation relations to consider and compare every execution path with the specification. 
$\mathcal{T}$ by-simulates a workload $\mathbf{Spec}$ if \textit{every} path of $\mathcal{T}$ (by-)simulates  $\mathbf{Spec}$.  
That is, every possible reconfiguration path in $\mathcal{T}$ meets all deadlines when executing the specified actions.

The scheme proposed by Pal et al., \cite{PalSPP16}, permits reconfiguration from $r$ to a weaker configuration $r'$ only when sufficient slack has been earned to permit a slower execution of the next action plus time for reconfiguration (before and possibly after),  i.e.,
$\langle r, t_{i-1} \rangle \trans{a_i}{w_i}
\langle r', t_{i} \rangle$ if $d_i - t_{i-1} \geq  2*\delta + \tau(r',a_i)$,
where $t_{i} = t_{i-1} + \delta + \tau(r',a_i)$ (i.e., $w_i = \delta_{r,r'} + \tau(r',a_i)$).
That is, the sum of the slack earned so far and the budgeted time for $a_i$ should exceed the time for reconfiguring and executing on a slower configuration, with a further allowance for a possible reconfiguration to a faster configuration to avoid missing future deadlines.
Otherwise, a faster configuration ($r_{max}$, to be safe) is chosen.
It thus defines a non-deterministic transition system $\mathcal{P}$ which is a subtransition system of the transition system $\mathcal{N}$ mentioned above.
Theorem \ref{CORRECTNESS} states the correctness of this scheme (and so of any deterministic algorithm based on it).

\begin{theorem}\label{CORRECTNESS}
If $r_{max}$ can execute each action $a_i$ of a workload within its corresponding \textit{budgeted time} $b_i$, then the  scheme of Pal et al. defines a transition system $\mathcal{P}$ that by-simulates $\mathbf{Spec}$.
\end{theorem}
Note that we have been able to state a general proof of the correctness of the scheme in the abstract, without positing any model relating configurations to speeds, and without any bounds on the times for any task in $\mathcal{A}$.  
Note also that the scheme does not consider idling between actions, since that would be counter-productive to meeting deadlines. 

\section{Better: Comparing Resources Based on Energy Efficiency}\label{ENERGY}

The motivation for dynamic reconfiguration is to save energy, since weaker configurations  consume less energy, providing an opportunity to trade off time versus energy cost. 
We  focus on amortising total energy consumption, subject to the constraint of meeting all deadlines (other objectives can also be formulated). 
Accordingly, we modify the transition system to have weights that also consider cumulative energy costs. 
We assume that energy costs for an action are given by a function  $\gamma(r, a)$, again making the simplifying assumption that energy costs are  determined only by the configuration $r$ and the action $a$. 
Let the energy cost of reconfiguration from $r$ to $r'$ be denoted $\theta_{r,r'}$ which for simplicity we assume to be 0 when $r=r'$ and a constant $\theta$ otherwise. 

The reconfigurable energy-aware transition system $\mathcal{E}$ for executing workloads can be modelled with 
$Q = \mathit{Conf} \times \R \times \R$;
 $O(\langle r, t, e \rangle)  = e$; 
and $\langle r, t, e \rangle \trans{a}{w}
\langle r', t', e' \rangle$ if $t' = t + \delta_{r,r'} + \tau(r', a)$ and
$e' = e + \theta_{r,r'} + \gamma(r', a) $.  
The initial state is $\langle r, 0, 0 \rangle$.
In the general setting, the weight domain can be seen as a composite monoid. 

Consider a workload $\mathbf{Spec}$ and two paths of $\pi, \pi'$ of $\mathcal{E}$ that  \textit{both} by-simulate $\mathbf{Spec}$.  We say that 
 $\pi$  is \textit{more efficient} than $\pi'$ if  $\pi$ simulates $\pi'$. 
 That is, $\pi$ does whatever actions $\pi'$ can (within the deadlines), but at lower cumulative energy cost at each step. 
 
The notion can be extended to transition systems $\mathcal{P}$ and $\mathcal{P}'$ that both by-simulate $\mathbf{Spec}$.
$\mathcal{P}$ is \textit{more efficient} than $\mathcal{P}'$ in executing $\mathbf{Spec}$ if for every execution path $\pi'$ of  $\mathcal{P}'$, there exists a path $\pi$ of  $\mathcal{P}$ such that $\pi$  is \textit{more efficient} than $\pi'$.  
This is a simulation relation between the transition systems.  

Note that a simulation relation allows $\mathcal{P}$ to contain paths that are \textit{not} more efficient than any path in $\mathcal{P}'$.
We therefore modify the notion of simulation to yield that of a \textit{betterment}:
\begin{definition}\label{BETTERMENT}
	Suppose
	$\mathcal{T}_1 = (Q_1, \mathcal{A}, \mathcal{W}, \xrightarrow{}, Q_{1o}, O_1)$ and 
	$\mathcal{T}_2 = (Q_2, \mathcal{A}, \mathcal{W}, \xrightarrow{}, Q_{2o},
	O_2)$ are cumulative weighted transition systems on the same input alphabet $\mathcal{A}$ and weight domain $\mathcal{W}$.
	A betterment relation between $\mathcal{T}_1$ and $\mathcal{T}_2$ 
	is a binary relation $R \subseteq Q_1 \times Q_2$ 
	such that $(p,q) \in R$ implies
	(i) $O_2(q) \leq_{\mathcal{W}} O_1(p)$; 
    and (ii) whenever $p \trans{a}{-} p'$ then there exists at least one $q'$ such that $q \trans{a}{-} q'$, and  for {\em every} $q'$ such that $q \trans{a}{-} q'$,  $(p',q') \in R$.
\end{definition}
We say $q$ \textit{betters} $p$ if $(p,q)$ is in \textit{some} betterment relation.  Transition system $\mathcal{T}_2$ betters $\mathcal{T}_1$ if for all $p \in Q_{1o}$, and \textit{every} $q \in Q_{2o}$, $(p,q)$ is in a betterment relation.
That is, every path in $\mathcal{T}_2$ is at least as efficient as any path in $\mathcal{T}_1$.  
In other words, $\mathcal{T}_2$ is in ``every way better'' than $\mathcal{T}_1$. 
Note that if $\mathcal{T}_2$ is deterministic, a betterment reduces to a simulation. 

The identity relation on transition systems may not be a betterment. 
However, betterments are closed under composition and union. 
\begin{proposition}\label{PROP-BETTERMENT}.
	(i) If $R_1, R_2$ are betterments, then so is $R_1 \circ R_2$. 
	(ii) If $R_i ~~(i \in I)$ are betterment relations between two given transition systems, then so is $\bigcup_{i \in I}~R_i$.
\end{proposition}

The scheme in \cite{PalSPP16} additionally examines the energy savings when opportunistically deciding to reconfigure,  i.e.,
$\langle r, t_{i-1}, e_{i-1} \rangle \trans{a_i}{e}
\langle r', t_{i}, e_{i} \rangle$ if
(i) $d_i - t_{i-1} \geq  2*\delta + \tau(r',a_i)$;
(ii) $\gamma(r,a_i) \geq \gamma(r',a_i) +  2 \theta$
where $t_{i} = t_{i-1} + \delta + \tau(r',a_i)$, and
$e_i =  e_{i-1} + e$, where $e = \theta_{r,r'} + \gamma(r',a_i)$.

\begin{theorem}\label{BETTER}
If baseline configuration $r_{max}$ can execute each action $a_i$ of a workload $\mathbf{Spec}$ within its corresponding \textit{budgeted time} $b_i$, then \textit{any} execution under the Pal et al. energy-saving scheme  \cite{PalSPP16} is a better (more efficient) by-simulation than execution on the baseline configuration $r_{max}$.
\end{theorem}

\section{What's Best?}\label{BEST}

The scheme in  \cite{PalSPP16} is \textit{not optimal} for arbitrary workloads.
For \textit{finite} workloads it is possible to determine optimal executions using offline techniques such as the YDS algorithm \cite{YaoDS}, or \textit{dynamic programming} techniques for related problems. 
However, it may not be pragmatic to use such offline algorithmic techniques because of the size of the workload and the available memory and computational resources.
Hence the problem is posed in a manner resembling an \textit{online algorithm} with an \textit{estimate} of the maximum time and energy required for executing the next action. 
However, one would like to ask how far from the optimal (either in absolute or relative terms) the approximation given by any given scheme is.  
We propose that simulation relations on cumulative weighted transition systems can provide a framework for reasoning about relative performance guarantees of approximations.
We extend the weight domain to being a semiring,
$\langle \mathcal{W}, \oplus, \oldstylenums{0}, \odot,  \oldstylenums{1} \rangle$, where $\oldstylenums{0}$ is the identity element for $\oplus$, and $\oldstylenums{1}$ is the identity element for $\odot$.

\begin{definition}\label{CONST-FACTOR}
	Suppose
	$\mathcal{T}_1 = (Q_1, \mathcal{A}, \mathcal{W}, \xrightarrow{}, Q_{1o}, O_1)$ and 
	$\mathcal{T}_2 = (Q_2, \mathcal{A}, \mathcal{W}, \xrightarrow{}, Q_{2o},
	O_2)$ are weighted transition systems on the same input alphabet $\mathcal{A}$ and weight domain $\mathcal{W}$.
	Let $c$ be any constant in $\mathcal{W}$.
	A constant-factor $c$-simulation relation between $\mathcal{T}_1$ and $\mathcal{T}_2$ 
	is a binary relation $R_c \subseteq Q_1 \times Q_2$ 
	such that  $(p,q) \in R_c$ implies
	(i) $O_2(q) \leq_{\mathcal{W}} c \odot O_1(p)$; 
	and (ii) whenever $p \trans{a}{-} p'$, there exists $q'$ such that $q \trans{a}{-} q'$ and $(p',q') \in R_c$.
\end{definition}

Constant factor simulations include the identity relation and are closed under relational composition  (which corresponds to $\odot$ on the indexing constants). Moreover they are monotone increasing with respect to the indexing constant. 
For any $c$, $c$-simulations are closed under union. 

\begin{proposition}\label{CONST-SIM}
	(i) The identity relation $\{ (p,p) ~|~ p \in Q \}$ is a $\oldstylenums{1}$-factor simulation;
	(ii) If $(p,q) \in R_c$ and $(q,s) \in R_{c'}$, then $(p,s) \in R_{c \odot c'}$
	(iii) If $c \leq c'$ then if $q$ can simulate $p$ up to constant factor $c$, then so can it up to constant factor $c'$.
	(iv) If $R_i~~(i \in I)$ are all $c$-simulations, then $\bigcup_{i \in I} R_i$ is also a $c$-simulation. 
\end{proposition}
An algorithm $A_2$ has a competitive ratio of $c$ with respect to another
$A_1$ if there is a $c$-simulation between the (deterministic) transition systems defined by them on any given input sequence of actions. 
That competitive ratio $c$ between two algorithms is \textit{tight} if there is no $c'$-simulation between them for any $c' < c$.
Note that if $\alpha$ is the ratio of the speeds between the fastest and slowest configurations, then the scheme of Pal \textit{et al.} will be $\alpha$-competitive. 
This is however a weak bound. 

\section{Conclusions}\label{CONCLUSIONS}

Inspired by practical problems encountered in multicore architectures,
we have presented an operational formulation of a general problem that involves finding feasible executions of a series of actions each to be completed within hard budgetary constraints (deadlines), and then comparing the cost of the feasible executions. 
There are several trade-offs that can be explored once the problem is amenable to an operational framework. 
While finite instances of such problems may be optimally solved ``offline'', using techniques such as dynamic programming or automata-based programming techniques, we pose the general problem in an online form, allowing for infinite executions, and unbounded state and data spaces. 
Such a formulation allows us to extend well-studied notions in concurrency theory such as simulation relations to the class of weighted transition systems, and thence to a general notion of algorithmic correctness and efficiency. 
The quantitative and timing aspects of the problem have motivated the use of interesting algebraic structures such as cumulative monoids.  
Typically semirings (e.g., a \textit{min}-+ algebra, also called a tropical semiring) are employed for formulating and comparing the behaviour of systems, especially in optimisation problems.

\paragraph{Other applications.}
To illustrate that our formulation is not merely about meeting deadlines and that it is not confined to video decoding, let us consider another application involving multicore machines, this time related to security.
Consider the problem of executing a series of actions each to be completed within a prescribed energy budget. 
Such problems are increasingly important in energy-oriented compiler optimisations. 
It is by now well established that an attacker can gain side-channel information about a computation by observing the power consumption characteristics of a  machine performing a computation \cite{KocherJJ}.
Such attacks exploit information leakage from mobile devices (smartphones, wireless payment devices etc.) that are widely used today.
Therefore we have the additional objective of minimising information leakage through this ``side channel''.
A common approach to thwarting the attacker's capability involves generating noise to obfuscate the power-consumption profile of the actions (instruction/job/task). 
The noise generator can be run on another core in parallel with the main computation, but this is at the cost of extra power consumption. 
Amortising energy consumption across the actions, we can minimise the leakage of power-profile based information from a subsequence of actions (using any energy credit earned when performing earlier tasks well within their budgeted energy). 

\paragraph{Concurrent actions.}
Our formulation involved resource allocation rather than task scheduling, since the problem was presented as a \textit{sequence of (atomic) actions} to be executed --- only one task is enabled at a time.
However, our problem finds an obvious generalisation that involves scheduling as  well, when we are presented with a \textit{sequence of  sets of actions} where each set of actions must be concurrently executed. 
At each step the set of concurrently-enabled actions are to be executed within their given deadlines. 
If the deadlines can be met by an interleaving of the atomic actions, then one can allocate a minimal set of required cores, thus minimising energy consumption while meeting all deadlines.
A scheduler tries to find such an interleaving for the set of concurrently enabled actions. 
In case two or more concurrent actions must be mutually exclusively executed,  they are suitably interleaved in a feasible schedule (if one exists).
Similar conditions apply if one task has to be executed in preference to another. 
Otherwise, if the set of concurrent actions cannot be interleaved, the scheduler tries to allocate disjoint sets of cores for the parallel execution of the actions, in a manner that minimises energy consumption while still meeting the deadlines of each task.
In these cases, we may additionally need to consider the costs of allocating cores and assigning tasks, as well as idling costs when tasks wait at a synchronisation point.
Note that the scheduler needs to work online, in that the particulars of sets of actions that will materialise in the future are not known to it. 

\paragraph{Future work.}
To our knowledge, the connections between algorithmic efficiency and  performance guarantees on the one hand, and operational formulations such as simulations and bisimulations on the other have not been adequately explored. 
We recently became aware of a particular subclass of problems for which this connection has been well formulated, namely the connection proposed by Aminof \textit{et al.}  between weighted \textit{finite-state automata} and online algorithms \cite{AminofKL}.
Their main insight is to relate the ``unbounded look ahead'' of optimal offline algorithms with nondeterminism, and  the ``no look ahead'' of online algorithms with determinism.
Our proposed relationship can be seen as an extension from finite state automata to general transition systems, replacing language equality with relations such as simulations (and bisimulations and prebisimulations).

We are currently looking at formulating and analysing online algorithms that may have better competitive ratios for the general energy minimisation problem, using, e.g., branch and bound techniques, etc., with the intention of proving tighter bounds. 
We are considering the cases where there is a limit on how far ahead one can execute actions (because of, say,  a bounded buffer for decoded frames) and when the online algorithm can look ahead at the characteristics of the next $k$ frames when deciding what configuration to choose. 
In the future, we would also like to examine the connections between \textit{absolute} performance guarantees and the framework of amortised bisimulations \cite{KiehnA-amortized}.


\bibliographystyle{eptcs}
\bibliography{BestBy.bib}

\end{document}